\documentclass[aps,pra,floatfix,twocolumn,nofootinbib]{revtex4}

\usepackage{amsmath}
\usepackage{bm}
\usepackage{euscript}
\usepackage{graphicx}
\usepackage{dcolumn}

\graphicspath{{./Figures/}}

\def\deg{\mbox{$^{\circ}$}}

\def\ea0{\mbox{ $ea_0$}}

\def\mF0{\mbox{ $\mu\Phi_0$}}
\def\mF0rtHz{\mbox{ $\mu\Phi_0/\sqrt{\rm Hz}$}}
\def\rtHz{\mbox{$\sqrt{\rm Hz}$}}

\def\ecm{\mbox{ e$\cdot$cm}}

\def\sE{{\ensuremath{\EuScript E}}}

\begin{document}

\setcounter{footnote}{0}

\title{The prospects for a new search for the electron electric dipole moment in solid Gadolinium iron garnet ceramics}

\author{A. O. Sushkov}
\email{alex.sushkov@yale.edu}
\affiliation{Yale University, Department of Physics, P.O. Box
208120, New Haven, CT 06520-8120}

\author{S. Eckel}
\affiliation{Yale University, Department of Physics, P.O. Box
208120, New Haven, CT 06520-8120}

\author{S. K. Lamoreaux}
\affiliation{Yale University, Department of Physics, P.O. Box
208120, New Haven, CT 06520-8120}

\date{\today}

\begin{abstract}
We address a number of issues regarding  solid state electron electric dipole moment (EDM) experiments, focusing on gadolinium iron garnet (abbreviated GdIG, chemical formula Gd$_3$Fe$_5$O$_{12}$) as a possible sample material. GdIG maintains its high magnetic susceptibility down to 4.2~K, which enhances the EDM-induced magnetization of a sample placed in an electric field. We estimate that lattice polarizability gives rise to an EDM enhancement factor of approximately 20. We also calculate the effect of the demagnetizing field for various sample geometries and permeabilities. Measurements of intrinsic GdIG magnetization noise are presented, and the fluctuation-dissipation theorem is used to compare our data with the measurements of the imaginary part of GdIG permeability at 4.2~K, showing good agreement above frequencies of a few hertz. We also observe how the demagnetizing field suppresses the noise-induced magnetic flux, confirming our calculations. The statistical sensitivity of an EDM search based on a solid GdIG sample is estimated to be on the same level as the present experimental limit. Such a measurement would be valuable, given the completely different methods and systematics involved. The most significant systematics in such an experiment are the magnetic hysteresis and the magneto-electric effect. Our analysis shows that it should be possible to control these at the level of statistical sensitivity.

\end{abstract}
 \maketitle

\section{Introduction}

One of the ways to test the Standard Model in low-energy experiments
is to search for parity and time-reversal-violating permanent
electric dipole moments of particles such as the electron and the
neutron. It is well-known that the observed matter-antimatter
asymmetry of the Universe can not be explained by the CP-violation
mechanisms within the Standard Model~\cite{Balazs2005}. A variety of theories suggesting new sources of CP-violation exist, many of them are already constrained by the current experimental limit on the electron EDM: $d_e<1.6\times 10^{-27}\ecm$~\cite{Regan2002}. A number of experimental EDM searches are currently under way, or being developed, they involve diatomic molecules~\cite{Hudson2002,Kawall2004,Kawall2005}, molecular ions~\cite{Stutz2004}, cold atoms~\cite{Weiss2003}, liquids~\cite{Ledbetter2005}, and solid-state systems~\cite{Bouchard2008}.

\begin{figure}[h!]
    \includegraphics[width=\columnwidth]{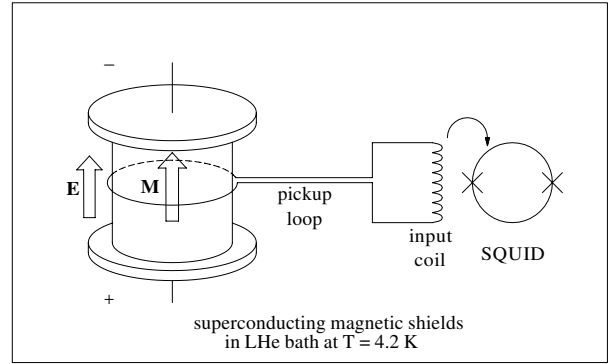}
    \caption{A schematic experimental setup for an EDM search.}
    \label{fig:EDMSetUp}
\end{figure}

Following the original suggestion of Shapiro \cite{Shapiro1968}, we
have been investigating the possibility of an improved limit on the
permanent electric dipole moment (EDM) of the electron by use of
paramagnetic insulating solids together with modern magnetometry
\cite{Lamoreaux2002}.  The basic idea is that when a paramagnetic
insulating solid sample is subjected to an electric field, if the
constituent atoms or ions of the solid have an electric dipole
moment, the atom or ions spins will tend to become spin-polarized.
Because the atoms or ions carry a magnetic moment, the sample
acquires a net magnetization, which can be detected by a SQUID magnetometer
inside a set of superconducting magnetic shields in a liquid helium bath
(see Fig.~\ref{fig:EDMSetUp}).

Three solid state electron EDM experiments have been reported in
literature. The first of these used a nickel-zinc ferrite, obtaining
a limit of about $10^{-22}\ecm$~\cite{Vasilev1978a}. Another
experiment, described in Ref.~\cite{Heidenreich2005}, obtained a
limit that is lower by a factor of 40. This experiment employed a
converse effect, in which a sample electric polarization appears
when the sample is magnetized. The third experiment
employs Gadolinium Gallium Garnet (GdGG), which has previously been
identified as a promising paramagnetic system for this type of
experiment~\cite{Liu2004}. We have performed preliminary
measurements, indicating that the anticipated sensitivity can be
achieved, pending a $1/T$ behavior of the paramagnetic
susceptibility to temperatures of order 10~mK. However, attaining an
improved electron EDM limit requires that GdGG remains a simple
paramagnetic system and that spin-glass or other effects do not
enter down to temperatures of order 10~mK. Indeed, it is now known that
such effects become very important near 0.2~K~\cite{Schiffer1995},
where the magnetic susceptibility reaches a peak value of $\chi=0.15$ (CGS units).
This value is actually larger than the susceptibility given by the Curie-Weiss
law, $\chi\propto (T-T_c)^{-1}$, which describes the susceptibility
of GdGG above 1~K, where $T_c\approx -2$~K for GdGG. We are
continuing to investigate the ultimate sensitivity of a GdGG-based
experiment, and the relaxation time of the spins remains to be
determined.
It is possible that this time will be long enough that
extra noise due to spin fluctuations will be a limitation
\cite{Budker2006}, as we discuss below. In the meantime we have
mounted a search for other materials whose properties make them
attractive for conducting this type of an EDM experiment.

\section{The search for the best material}

\subsection{The magnetic susceptibility}

For a solid state electron EDM experiment, materials with large
magnetic susceptibility (or relative permeability) are desired. This
is because, neglecting boundary effects (see Section~\ref{sec:g}), the EDM-induced magnetization is given by
\begin{equation}\label{equ:M1}
M = \chi k d_e E/\mu_a,
\end{equation}
where $\chi$ is the magnetic susceptibility, $k$ is the effective
EDM enhancement factor in the material under study, $d_e$ is the
electron EDM, $E$ is the applied electric field, and $\mu_a$ is the
magnetic moment of the paramagnetic atoms or ions that carry the
electron EDM ($\mu_a=8\mu_B$ for Gd$^{+3}$, $\mu_B$ is the Bohr magneton).
This assumes that all of the
magnetization is due to the magnetic moments of the heavy atoms, that carry an EDM (a good
approximation for all cases considered here). We see immediately
that $\chi$ should be as large as possible. Note that the sample
temperature does not explicitly enter into this formula.

Garnet ferrites with rare earth substitution can have large
permeabilities at temperatures below 100 K. We have studied the
complex permeability ($\mu=\mu'-i\mu''$, $\mu = 4\pi\chi$) of mixed
Gadolinium-Yittrium iron garnets (GdYIG, chemical formula
Gd$_{(3-x)}$Y$_x$Fe$_5$O$_{12}$) down to 2~K~\cite{Eckel2008}. Our
data show that for pure GdIG ceramic, $\mu' = 77$ at 4.2~K, and for
mixed Gd$_{1.8}$Y$_{1.2}$Fe$_5$O$_{12}$ ceramic, $\mu' = 50$ at 4.2~K.
The immediate conclusion is that at 4.2~K, in the absence of
sample-shape effects, the EDM-induced magnetization in pure GdIG is
enhanced by a factor of more than 300 compared to that in GdGG, which has $\chi = 0.016$ at 4.2~K~\cite{Schiffer1995}. The present
manuscript describes our study of the feasibility of an EDM
experiment with GdIG.

\subsection{The effective EDM enhancement factor and the internal
electric field}

The enhancement factor $k$ that appears in Eq.~(\ref{equ:M1}) determines the
scale of the effective electric field acting on the Gd$^{+3}$ ion
EDM in GdIG. To calculate this, let us consider the energy shift
$\delta$ resulting from a non-zero EDM of an ion in the lattice of
an ionic solid. A naive estimate of this energy shift can be written
down as follows:
\begin{equation}
\label{equ:naive1} \delta_{\text{naive}} \approx -d_i
E_{\text{local}} \approx -\frac{\epsilon+2}{3}d_iE,
\end{equation}
where $d_i$ is the Gd$^{+3}$ ion EDM, $E_{\text{local}}$ is the
local electric field at a Gd$^{+3}$ ion site, $\epsilon$ is the
dielectric constant of GdIG, and $E$ is the applied electric field.
Here we used the Lorentz relation for the local field in a
dielectric (in the approximation of a cubic crystal structure)~\cite{Kittel}. The magnitude of $d_i$ induced by the
electron EDM is calculated in Ref.~\cite{Dzuba2002}: $|d_i| \approx 2.2d_e$\footnote{This naive estimate of the enhancement factor gives the incorrect sign if the actual value $d_i \approx -2.2d_e$ from Ref.~\cite{Dzuba2002} is used. The issue of this sign is addressed in Ref.~\cite{Mukhamedjanov2003}}.
With $\epsilon=15$, we get the naive estimate $\delta_{\text{naive}}
\approx -13 d_eE$.

A more rigorous calculation is presented in
Refs.~\cite{Mukhamedjanov2003,Dzuba2002,Kuenzi2002}. The energy shift is
expressed in terms of the displacement $x$ of the EDM-carrying ion
with respect to its equilibrium position in the unit cell:
\begin{equation}
\label{equ:rig1} \delta \approx
-0.1\frac{x}{a_B}\frac{d_e}{ea_B}\cdot 27.2\text{ eV},
\end{equation}
where $a_B$ is the Bohr radius, $e$ is electron charge, and
$27.2\text{ eV}$ is the atomic unit of energy. The Gd$^{+3}$ ion
displacement $x$ is related to the applied electric field through
the dielectric constant: $x=(\epsilon_{ion}-1)E/12\pi e n$, where
$\epsilon_{ion}$ is the dielectric constant contribution due to ion
displacement and $n\approx 10^{22}$~cm$^{-3}$ is the Gd density. We
are assuming that the displacement of Fe ions is much smaller due to
the stiffer Fe-O bond. The other contribution to the dielectric
constant in GdIG is the ``electronic polarizability'' - the
polarization of the individual ions in the electric field. We can
estimate this contribution from the refractive index $n\approx 2.2$
of GdIG at optical frequencies, where the lattice polarizability
does not contribute. Thus $\epsilon_{ion}\approx 15 - 2.2^2 \approx 10$ is a
reasonable estimate for the lattice polarizability. With this value,
the EDM-induced energy shift in GdIG is:
\begin{equation}
\label{equ:final} \delta \approx -20 d_e E.
\end{equation}
Thus the effective enhancement factor is $k = 20$. As
pointed out in Ref.~\cite{Dzuba2002}, this is a conservative
estimate, and the true result may be up to a factor of two larger.
Note that the final value of $k$ is quite close to the naive
estimate discussed above.

\section{The effect of the sample shape}\label{sec:g}

Above we described two enhancement mechanisms for an EDM-induced
signal in GdYIG ferrites: large magnetic permeability and local
electric field enhancement by the crystal lattice. The combination
of these makes this an attractive system for mounting an EDM search.
It must be noted, however, that sample geometry-dependent
demagnetizing fields eliminate part of the advantage of having a
large-permeability material. It is well known, for example, that it
is difficult to magnetize a thin disk of large-permeability magnetic material along its
axis of symmetry (permanent magnets with such magnetization exist because their magnetization is saturated, i.e. their linear permeability is close to 1). This is explained by the Maxwell's equation
$\nabla\cdot\bm{B} = 0$, combined with the relation $\bm{B} = \bm{H}
+ 4\pi\bm{M}$ (CGS units). If a sample has some macroscopic
magnetization $\bm{M}$, then on its surface $\nabla\cdot\bm{H} =
-4\pi\nabla\cdot\bm{M} \neq 0$ gives rise to the ``demagnetizing
field'' $\bm{H}$, which opposes the magnetization $\bm{M}$. The
magnitude of this demagnetizing field depends on the sample
geometry and permeability. For a high-permeability material the resultant
magnetization can be much less than that in the case of no
demagnetizing field, which happens for a long thin needle geometry,
for example.

\begin{figure}[h!]
    \includegraphics[width=\columnwidth]{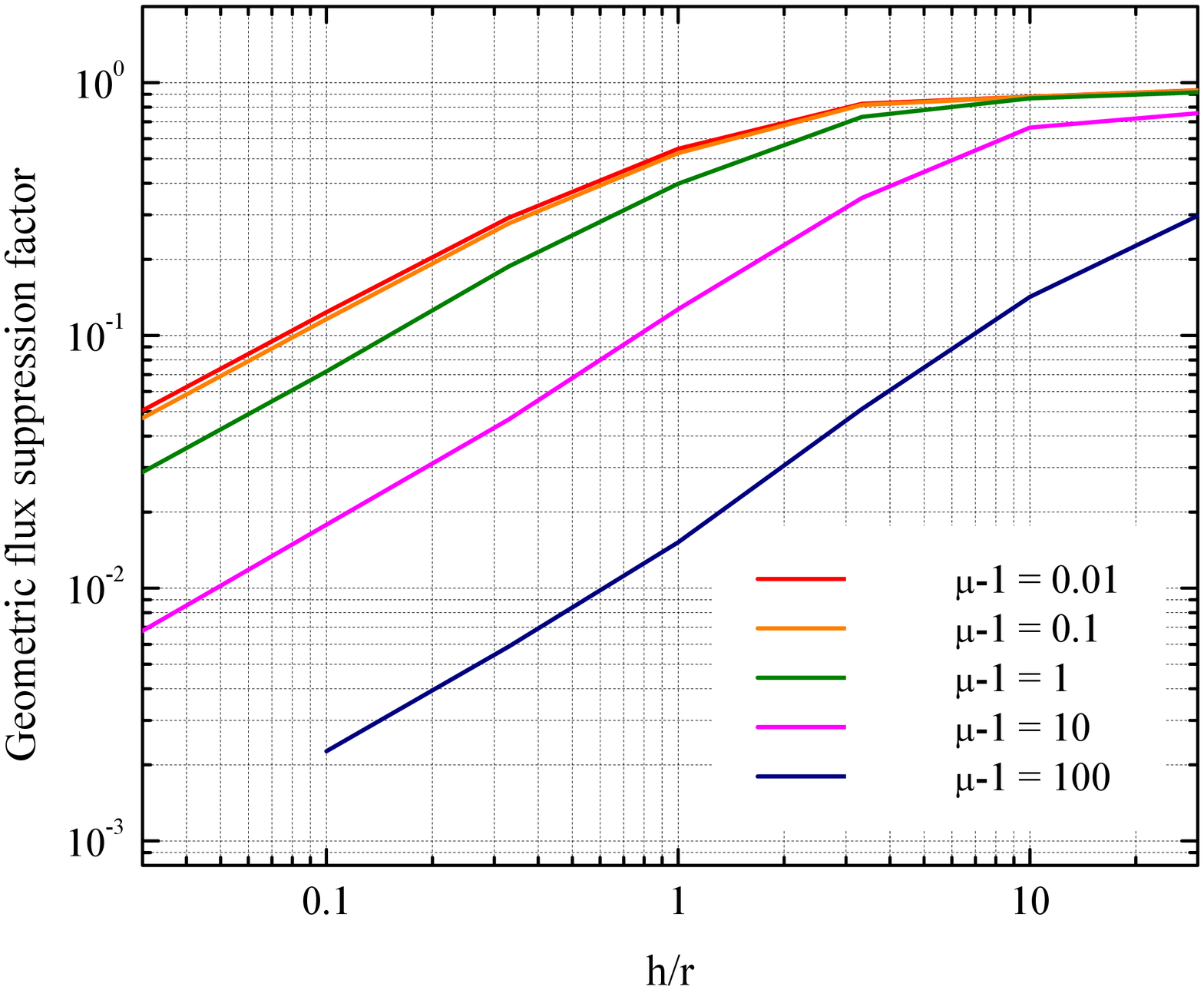}
    \caption{The magnetic flux suppression geometric factor for a permeable cylindrical sample.}
    \label{fig:GF}
\end{figure}

The demagnetizing field, and the sample magnetization, can be
calculated exactly when the material is in the form of an
ellipsoid~\cite{Landau8}, but not for a disk-shaped sample, which
would be used an an EDM search (some approximate expressions for saturated hard ferromagnets are obtained in Refs.~\cite{Joseph1965,Joseph1966,Arrott1979,Sato1989}). We used the finite-difference method
to numerically solve Maxwell's equations $\nabla\cdot\bm{B} = 0$ and
$\nabla\times\bm{H} = 0$, with a cylindrical sample of radius $r$,
thickness $h$, and permeability $\mu$. In such a geometry the
magnetization is not uniform, but we can calculate the total
magnetic flux $\Phi$ through a circular loop positioned on the
surface of the cylinder halfway between its top and bottom. When
$h\gg r$, there is no demagnetizing field, and $\Phi = 4\pi M \pi
r^2$, where $M$ is the EDM-induced magnetization given by
Eq.~(\ref{equ:M1}). In general, however, we have to write
\begin{equation}
\label{equ:Phi1} \Phi = g\cdot4\pi M \pi r^2,
\end{equation}
where $g = g(h/r,\mu)$ is the flux suppression geometric factor that depends
of the geometry and the permeability of the sample. The results of
our numerical calculations, expressed as $g(h/r,\mu)$, are shown in
Fig.~\ref{fig:GF}.

There is clearly a significant suppression of the EDM-induced
magnetic flux for a reasonable sample geometry ($h/r \simeq 1$) and
for $\mu\simeq 50$. Certainly the suppression factor $g$ does not
give the whole story, since there are several other
geometry-dependent factors that enter into the final EDM
sensitivity, such as the pickup loop area and inductance, as well as
the electric field for a given experimentally achievable voltage
applied to the sample. Further discussion of these factors
can be found in Section~\ref{sec:sensitivity}. It is also important to note that
for $h/r \ll 1$, there is a significant suppression even for
$\mu-1\ll 1$, as would be the case in a search for a nuclear Schiff
moment, for example~\cite{Bouchard2008}. Therefore, in any
magnetization-type EDM search, a thin disk-shaped sample would lead
to a loss in sensitivity, instead samples with $h \simeq r$ must be
used\footnote{This conclusion is also valid if an atomic magnetometer (or some other magnetometer) is used to detect the magnetic field outside the sample, instead of a SQUID that detects the magnetic flux through a pickup loop.}.

\section{The magnetic noise measurements}

\begin{figure}[b!]
    \includegraphics[width=\columnwidth]{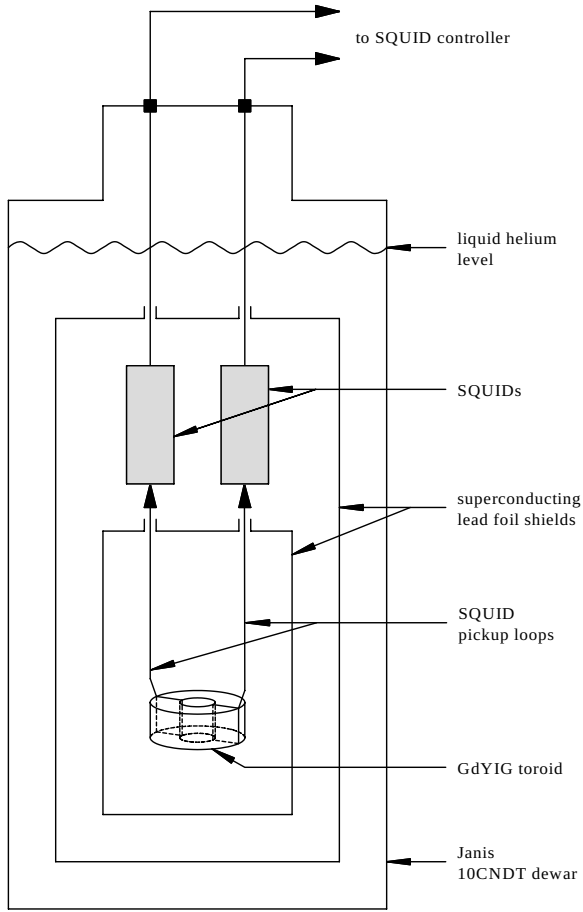}
    \caption{The experimental setup for magnetic noise measurements.}
    \label{fig:SetUp}
\end{figure}

Even with a substantial signal suppression by the demagnetizing
fields, it is possible, by carefully choosing the sample dimensions, to design an EDM search of formidable
sensitivity. Before a full-scale experiment with Gd-containing
ferrites is attempted, however, the
intrinsic magnetic noise of the material must be measured. If this noise is greater
than the magnetometer noise, the sensitivity of the EDM-induced
magnetization measurement is reduced~\cite{Budker2006}.

\begin{figure}[b!]
    \includegraphics[width=\columnwidth]{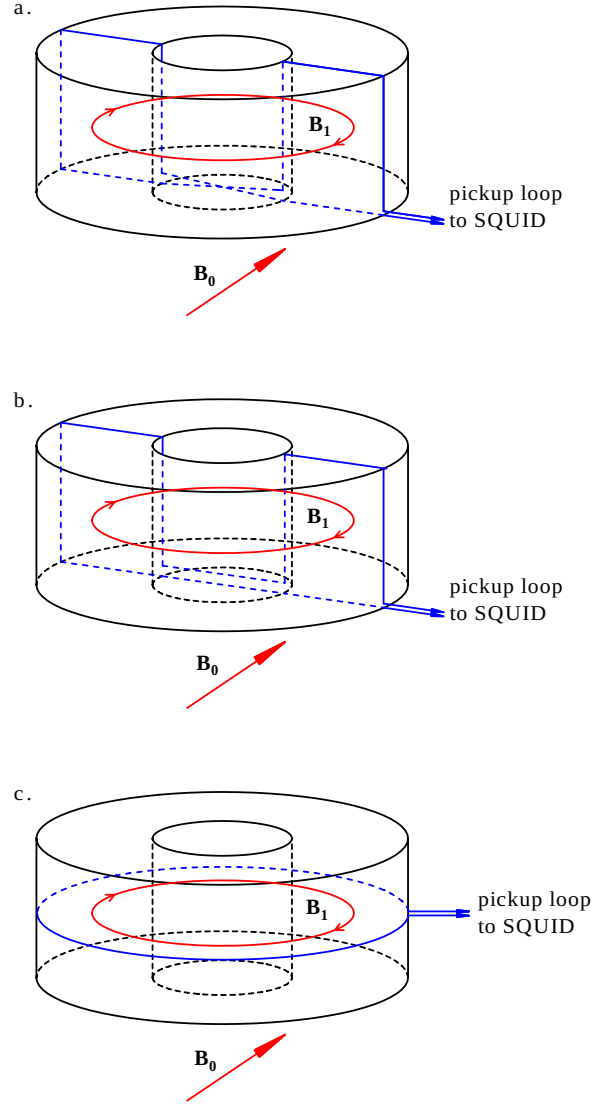}
    \caption{The three SQUID pickup loop geometries used for the magnetic noise measurements:
    a. pickup loop in the shape of figure-eight for external magnetic field rejection;
    b. U-shaped pickup loop for rejection of magnetic field lines internal to the toroid;
    c. pickup loop wound around the toroid perimeter to detect axial magnetic field.
    Magnetic field vectors are schematically shown in red. $B_0$ is the magnetic field outside the toroid. $B_1$ labels an azimuthal magnetic field line, looping around inside the toroid.}
    \label{fig:pickup}
\end{figure}
\begin{figure}[b!]
    \includegraphics[width=\columnwidth]{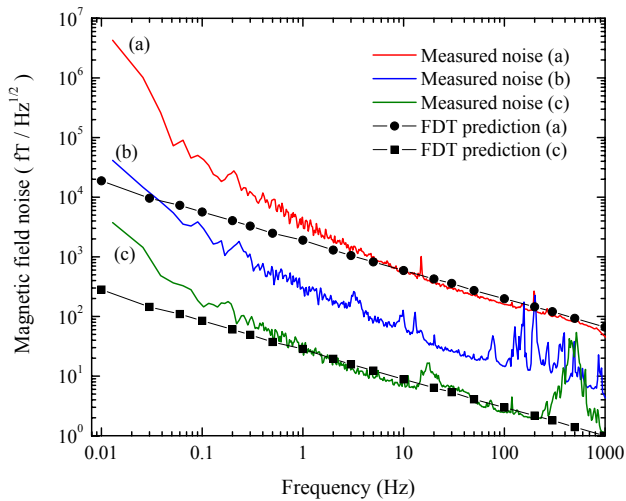}
    \caption{The results of the magnetic noise measurements.
    The curves are marked with the corresponding pickup loop geometry, shown in Fig.~\ref{fig:pickup}.
    Two noise expected from the fluctuation-dissipation theorem (FDT) for geometries (a) and (c) is also shown.}
    \label{fig:Noise1}
\end{figure}

The setup used for the GdIG and GdYIG magnetic noise measurements is shown in Fig.~\ref{fig:SetUp}. The sample is a sintered
ceramic toroid with a square cross-section, outer diameter 3~cm, inner diameter 1~cm, and thickness 1~cm. Experiments were
performed with ceramic samples of Gd$_3$Fe$_5$O$_{12}$, obtained
from Pacific Ceramics~\cite{PacificCeramics}. The superconducting pickup loop was made from enamel-insulated niobium wire of
thickness 0.003~in. It was connected to the input terminals of a Quantum Design DC thin-film SQUID sensor (model 50). The SQUIDs were connected to the control unit (model 5000) via 4-meter MicroPREAMP cables. The measurement bandwidth was limited to 1~kHz by an analog filter in the SQUID controller, to prevent aliasing. The intrinsic flux noise of the SQUID magnetometers in this frequency range was 2~$\mu\Phi_0$/\rtHz, which corresponds to magnetic field noise of several fT/\rtHz, the exact value depending on the area of each pickup loop. This noise level was always much less than the measured sample magnetization noise. The SQUID noise 1/f corner was around 0.1~Hz.
The SQUIDs and the sample were mounted on G-10
holders, carefully fixed to minimize vibrations, and enclosed in superconducting shielding, made of 0.1-mm thick Pb foil, glued to the inner surfaces of two G-10
cylinders. The magnetic field shielding factor of our setup was greater than $10^9$.
The entire setup was submerged in liquid helium
inside a Janis model 10CNDT dewar. To dampen out vibrations, the dewar was mounted on rubber supports and placed on a 15" diameter support ring, comprising a partially inflated bicycle inner tube; however the residual vibrations of the SQUID pickup loops with the respect to the residual magnetic field gradients\footnote{These gradients are probably caused by the magnetic flux trapped at some pinholes in the Pb foil.} inside the superconducting magnetic shields were enough to cause the pronounced high-frequency (between 10~Hz and 1~kHz) peaks in the SQUID measurements shown in Fig.~\ref{fig:Noise1}.
Temperature was measured with a LakeShore model DT-670C-SD silicon diode
temperature sensor, accurate to 0.1~K.

Magnetic noise measurements were done with several pickup-loop geometries to investigate the effect of the demagnetizing
field on the noise. It is energetically favorable for the magnetic field lines to stay inside a high-permeability material, therefore most of the flux
due to the magnetic noise in the material is confined inside the toroid, as indicated in Fig.~\ref{fig:pickup}. This flux can be detected by a SQUID pickup loop wound in the form of a figure-eight, i.e. in the gradiometer configuration (Fig.~\ref{fig:pickup}a). The spectrum of the measured magnetic field noise for this
geometry is shown as the red curve marked (a) in Fig.~\ref{fig:Noise1}. In this case, the magnetic field lines were closed inside the toroid, and, to a good approximation, there was no demagnetizing field (geometric factor $g=1$), therefore this was a measurement of the intrinsic magnetization noise of the ferrite sample.

To demonstrate that the observed magnetic field noise is due to the magnetization inside the material, rather than the magnetic field noise originating outside the sample, we performed noise measurements with a SQUID pickup loop wound in the U-shaped form, i.e. in the magnetometer configuration perpendicular to the plane of the toroid (Fig.~\ref{fig:pickup}b). The spectrum of the measured magnetic field noise for this
geometry is shown as the green curve marked (b) in Fig.~\ref{fig:Noise1}. In this case, the flux due to the magnetic field lines that were closed inside the toroid was not coupled to the SQUID, and only the magnetic field that entered or exited the toroid was detected.

Finally, to simulate the geometry of an EDM experiment as closely as possible, we measured the magnetic field noise that couples to the SQUID with the pickup loop wound around the perimeter of the sample, as shown in Fig.~\ref{fig:pickup}c. In this case, the magnetic flux noise was given by the GdIG magnetization noise, reduced by the appropriate geometric factor that takes into account the demagnetizing field, as in Eq.~(\ref{equ:Phi1}). The results are shown as the blue curve marked (c) in Fig.~\ref{fig:Noise1}.
The pronounced peaks in all three of the noise data sets are caused by vibrations, as discussed above.

Let us proceed to the analysis and discussion of these magnetic noise data.
The magnetization noise in a material with a known complex permeability $\mu = \mu'-i\mu''$ can be calculated using the fluctuation-dissipation theorem (FDT)~\cite{Landau5}:
\begin{equation}
\label{equ:FDT} (M^2)_{\omega} = \frac{1}{2\pi V}\frac{k_BT}{\omega}\mu''(\omega),
\end{equation}
where $(M^2)_{\omega}$ is the power spectral density of the magnetization noise at frequency $\omega$, $V$ is the sample volume, $k_B$ is Boltzmann's constant, and $T$ is temperature. Complex permeability has been measured in GdIG and mixed GdYIG ceramics in a range of temperatures (2~K to 295~K) and frequencies (100~Hz to 200~MHz)~\cite{Eckel2008}. In order to compare our noise measurements with the FDT, we measured the complex permeability of GdIG at 4.2~K in the frequency range of 0.01~Hz to 1~kHz. This was done by winding two 23-turn excitation coils in series, one around the GdIG toroid, and one around a toroid of the same dimensions, but made of G-10. Each of the toroids was connected to a SQUID via a one-turn pickup loop. The GdIG complex permeability was extracted from the relative amplitude and phase shift of the SQUID signals, as a sinusoidal current with frequency ranging between 0.01~Hz and 1~kHz was applied to the excitation coil. The resulting values of $\mu''$ were used in Eq.~(\ref{equ:FDT}) to generate the magnetic field noise expected from the FDT. This is shown by circles in Fig.~\ref{fig:Noise1}, in the case of no demagnetizing fields (geometric factor $g=1$), which is a good approximating for geometry (a).

For frequencies greater than approximately 5~Hz, there is very good agreement with the magnetic field noise measured with the SQUID for this geometry. We interpret this as a verification of the validity of the FDT for our system in this frequency range. For lower frequencies, however, the FDT seems to systematically underestimate the magnetic field noise, by as much as two orders of magnitude. We speculate that the excess magnetic field noise is due to very slowly-relaxing degrees of freedom, which come to thermal equilibrium on the time scales much longer than the duration of our experimental run (a few hours)~\cite{Crisanti2003}. For such degrees of freedom, the effective noise temperature that should be used in Eq.~(\ref{equ:FDT}) is much greater than 4.2~K, thus they contribute greater noise.

In Section~\ref{sec:g} we discussed how the sample shape affects the EDM signal, presumably the magnetization noise is also reduced by the same demagnetizing fields. This is confirmed by the measurements of the noise with the SQUID pickup loop around the sample perimeter (Fig~\ref{fig:pickup}c). Indeed, if the FDT prediction (\ref{equ:FDT}) is multiplied by the geometric factor appropriate for the sample dimensions ($h/r = 0.67$ resulting in $g=0.017$)\footnote{We approximate the sample as a disk. Our simulations show that the presence of the central hole does not change the geometric factor by more than a few percent}, the result is the expected magnetic noise in the geometry of Fig~\ref{fig:pickup}c. This is shown by the squares in Fig~\ref{fig:Noise1}, and once again, the agreement with the SQUID measurements is remarkable above frequencies on the order of a few Hz.
We can therefore conclude that we predict the magnetic field noise in any geometry by taking into account the geometric factor, as described in Section~\ref{sec:g}.

\section{The projected sensitivity of an experiment with GdIG}\label{sec:sensitivity}

Having measured the magnetization noise in GdIG, we can estimate the statistical sensitivity of an EDM experiment with a GdIG sample. Combining Eqs.~(\ref{equ:M1}) and~(\ref{equ:Phi1}), we obtain the EDM-induced magnetic flux through a pickup loop around a cylindrical sample:
\begin{equation}
\label{equ:EDM2} \Phi_{\text{EDM}} = g\pi r^2\cdot4\pi \chi\frac{kd_eE}{8\mu_B},
\end{equation}
where all the symbols have been defined previously. The statistical sensitivity is limited by the material magnetization noise, given by Eq.~(\ref{equ:FDT}), the resulting magnetic flux noise through the pickup loop is given by:
\begin{equation}
\label{equ:EDM3} \Phi_{\text{noise}} = g\pi r^2\cdot4\pi \sqrt{\frac{1}{4\pi^3r^2h}\frac{k_BT\mu''}{f}},
\end{equation}
where $f=\omega/2\pi$ is the electric field reversal frequency. We can now express the statistical sensitivity in terms of experimental parameters:
\begin{equation}
\label{equ:EDM4} \delta d_e = \frac{4\mu_B}{\pi\chi kV_0}\sqrt{\frac{h}{\pi r^2}\frac{k_BT\mu''}{f}},
\end{equation}
where we introduced the amplitude $V_0$ of the voltage applied to the electrodes. Interestingly, the geometric factor $g$ does not enter this sensitivity estimate, this is because it appears in both the EDM-induced flux and the noise-induced flux. If the magnetization noise had turned out to be lower than the intrinsic noise of the SQUID, the geometric factor would have entered into the sensitivity estimate (so would the SQUID sensitivity and the number of turns of the SQUID pickup loop, see Appendix~\ref{sec:app1}). Equation~(\ref{equ:EDM4}) assumes a disc-shaped geometry, but the sample dimensions have not been specified. It is clear that the best sample is a thin disc of large radius: decreasing the thickness $h$ increases the applied electric field for a given voltage $V_0$, and increasing the radius $r$ increases the sample volume, which decreases the magnetization noise, according to Eq.~(\ref{equ:FDT}). It is also clearly advantageous to reverse the electric field as fast as possible, however displacement currents limit the maximum realistic reversal frequency to 10~Hz (see below for more detailed discussion of systematics). For a quantitative estimate of the EDM sensitivity we take the following experimental parameters: $V_0=10$~kV, $h=0.1$~cm, $r=3$~cm, $f=10$~Hz, resulting in $\delta d_e \simeq 4\times 10^{-24}\ecm/\rtHz$. After 10 days of integration, a statistical sensitivity at the level of
\begin{equation}
\label{equ:EDM5} \delta d_e \simeq 3\times 10^{-27}\ecm
\end{equation}
can be achieved.

A number of systematic concerns for a solid state sample-based EDM search have been outlined in Ref.~\cite{Lamoreaux2002}. With a
ferromagnetic system, magnetic hysteresis is also crucially important. The main concern is electric field-correlated sample magnetization caused by the magnetic fields generated by displacement currents. It should be noted that these magnetic fields tend to be perpendicular to the EDM-induced magnetization, but some degree of misalignment is inevitable.  The current associated with applying 10~kV to a 300~pF sample at 10~Hz reversal frequency is about 30~$\mu$A, which gives
rise to magnetic field of order $H_d\simeq 2\times 10^{-6}$~A/cm. The resulting remanent magnetization $M_d$, present after the current is brought back down to zero, can be calculated from the parameters of the Raleigh hysteresis loops measured for GdIG in Ref.~\cite{Eckel2008}. Taking a reduction of a factor of 10 due to perpendicularity, the resulting magnetization along the electric field is $M_d^{||}\simeq 10^{-12}$~emu/cm$^3$. According to Eq.~(\ref{equ:Phi1}), the induced magnetic flux is suppressed by the geometric factor, which, for a disc of $h=0.1$~cm and $r=3$~cm, is approximately $10^{-3}$. The resulting systematic is at the level of statistical sensitivity given in Eq.~(\ref{equ:EDM5}).

Another possible systematic is the so-called magneto-electric effect~\cite{Landau8}. The applied electric field induces a strain in the sample (electrostriction), which then induces a magnetization correlated with the electric field, provided the sample already possesses some non-zero magnetization (inverse magnetostriction). This effect is described by the term $\gamma E^2 H^2$ (we omit the tensor indices) in the electromagnetic free energy, which is allowed by both the parity and the time-reversal symmetries. This has been observed in YIG~\cite{ODell1967} and GdIG~\cite{Mercier1974}. The ``magneto-electric susceptibility'' $\beta_M$ is defined in the following manner: $M = \beta_m(H_0)E^2$, where $H_0$ is the external magnetic field, $E$ is the applied electric field, and $M$ is the resulting magnetization. For GdIG single crystals in the low-field limit ($H_0<30$~Oe), the magneto-electric susceptibility was measured to be: $\beta_m(H_0) \simeq 10^{-8}\cdot H_0(\text{Oe})$~emu/cm$^3$, where $H_0$ is expressed in Oestreds~\cite{Mercier1974}. For a poly-crystalline sample, this should be reduced by a factor of 3 (the average over crystallite orientations). With the electric field of 100~kV/cm, and assuming 1\% reversal accuracy, the external magnetic field $H_0$ has to be kept below $3\times 10^{-7}$~Oe in order to control this systematic at the level of statistical sensitivity. This appears feasible with a combination of Metglas magnetic shielding and superconducting lead shields.

Finally, we point out that, if a gradiometer setup is used with a large-permeability sample, the gradiometer tuning has to be adjusted, see Appendix~\ref{sec:app2} for analysis.

\section{Conclusion and outlook}

We have performed a detailed feasibility study for an electron electric dipole moment search with a solid ceramic sample of Gadolinium iron garnet. This material is attractive because it is an excellent insulator, and it maintains a high magnetic permeability down to liquid helium temperatures, $\mu = 77$ at 4.2~K, which enhances the EDM-induced magnetization and allows the design of an experiment based on SQUID-magnetometer detection and superconducting magnetic shielding. In addition, the electric field acting on the electron EDM is amplified by the crystalline lattice polarizability, resulting in an effective EDM enhancement factor of 20. When calculating the EDM-induced magnetic flux, one must take into account the demagnetizing field, which suppresses it by a factor that depends on the sample geometry and permeability. The dominant sensitivity limitation in the proposed experiment, however, is the intrinsic magnetization noise of the Gadolinium iron garnet samples at 4.2~K. We measured this noise and verified that our data are consistent with the fluctuation-dissipation theorem in the 5~Hz to 1~kHz frequency range. At lower frequencies the noise seems to be dominated by magnetic degrees of freedom that are out of equilibrium with the liquid helium heat bath. We also verified the expected scaling of the noise-induced magnetic flux with the sample geometry. We estimate the statistical sensitivity of an EDM search based on a solid GdIG sample to be $3\times 10^{-27}\ecm$ after 10 days of averaging. This is slightly above the present experimental limit, but such a measurement would still be highly valuable, given the completely different methods and systematics involved. We also estimate the most likely systematics: the magnetic hysteresis and the magneto-electric effect. Our analysis shows that it should be possible to control these at the level of statistical sensitivity.

It seems feasible that the magnetic noise can be reduced by diluting the GdIG powder with a non-magnetic mixer (such as teflon) prior to pressing the ceramic samples, Another possibility is substituting some of the Fe ions in the lattice with paramagnetic ions, such as Ga, making mixed Gd$_3$Fe$_{(5-x)}$Ga$_x$O$_{12}$ ceramics. This will probably reduce the magnetic susceptibility as well as the magnetization noise, but, by carefully choosing the sample dimensions, this loss can be compensated by an improved geometric factor. Yittrium-substituted Gadolinium garnets with chemical formula Gd$_{3-x}$Y$_x$Fe$_5$O$_{12}$ have been studied in Ref.~\cite{Eckel2008}, and the measured complex permeability at 4.2~K leads to a poorer EDM sensitivity estimate for such ceramics.
We have, in addition, identified an extremely promising material where the sensitivity gain is provided by the crystal lattice, rather than by the large magnetic permeability~\cite{Mukhamedjanov2005}. This material is gadolinium molybdate (chemical formula Gd$_2$(MoO$_4$)$_3$), which is a ferroelectric-ferroelastic with Curie temperature of 160$\deg$C and spontaneous polarization of 0.25~$\mu$C/cm$^2$. Preliminary estimates and measurements indicate that a two-order of magnitude improvement over the present limit is possible with this material. We are currently pursuing further studies of Gd$_2$(MoO$_4$)$_3$, as well as other magnetic ferroelectrics.

\section{Acknowledgements}

The authors acknowledge useful discussions with Woo-Joong Kim, David DeMille, Jack Harris, Oleg Sushkov, Dima Budker, Michel Devoret, and Larry Hunter.

\appendix

\section{The multi-turn SQUID pickup loop}\label{sec:app1}

If the sample magnetization noise happens to be below the SQUID noise, the EDM sensitivity is limited by the SQUID noise. To achieve the best coupling of the sample flux to the SQUID, a multi-turn pickup loop may be used. This appendix shows that having a several-turn pickup loop may be advantageous, and suggests how the number of turns may be optimized.

Let us denote by $N$ the number of turns of the SQUID superconducting pickup loop around the sample, $A$ is the area of the sample. The circuit is schematically shown in Fig.~\ref{fig:EDMSetUp}. Suppose the magnetic field through the pickup loop  varies sinusoidally as a function of time, with frequency $\omega$: $B = B_0e^{-i\omega t}$. The electromotive force around the circuit is then given by Faraday's law: $\sE = i\omega N A B_0e^{-i\omega t} = i\omega N \Phi$, where we
define the magnetic flux $\Phi = BA$. The current in the circuit is given by
$I = \sE/(Z_{\text{in}}+Z_p)$, where the impedances of the SQUID
input coil and the pickup loop are in the denominator.
Expressing these inductive impedances as $Z = i\omega L$ results in $I = N \Phi/(L_{\text{in}}+L_p^{(N)})$, where
$L_{\text{in}}\approx 1.5$~$\mu$H is the input coil self-inductance and
$L_p^{(N)}$ is the self-inductance of the N-turn pickup loop. The
flux that is coupled to the SQUID is given by $\Phi_{\text{SQ}} =
M_{\text{in}}I$, where $M_{\text{in}}\approx 10$~nH is the SQUID-input coil
mutual inductance. Thus we finally have:
\begin{equation}
\label{equ:PhiSQ2} \Phi_{\text{SQ}} = \frac{N
M_{\text{in}}}{L_{\text{in}}+L_p^{(N)}}\Phi.
\end{equation}
Having a several-turn pickup loop improves the flux coupling to the
SQUID, but only until the pickup coil inductance reaches $L_p^{(N)}\approx
L_{\text{in}}$. If the number of turns is increased further, so that
$L_p^{(N)}\gg L_{\text{in}}$, the coupling deteriorates, since $L_p^{(N)}$ scales as $N^2$. The optimal number of turns can be calculated from Eq.~(\ref{equ:PhiSQ2}), given the dimensions of the sample and the superconducting wire, and using an expression for the inductance of a multi-turn coil~\cite{Terman1943}.

\section{Detuning of the Gradiometer Condition due to Sample Permeability}
\label{sec:app2}

Previous work \cite{Liu2004} employed two superconducting pickup
loops around the sample, operated in a concentric loop planar
gradiometer configuration, to reduce magnetic field noise.
Unfortunately, the permeability of the sample alters the pickup loop
inductances, and spoils the gradiometer tuning, and reduces the
experiment sensitivity slightly because the loop inductance is
increased. In the case of a single loop or radius $a$ around an
infinitely long cylinder of permeability $\mu$ and radius $b$, the
increase in inductance is \cite{Smythe1950}
\begin{equation}
\delta L= 2\mu_0 a^2\int_0^\infty \Phi(k)[K_1(ka)]^2 dk
\end{equation}
where $\mu_0$ is the permeability of free space, $K_1(ka)$ is the
first order modified Bessel function, and
\begin{equation}
\Phi(k)= {(\mu -1)kb I_0(kb)I_1(kb)\over (\mu -1)kb
K_0(kb)I_1(kb)+1}\approx {I_0(kb)\over K_0(kb)}
\end{equation}
where $I_0$ and $K_0$ are the zero-order modified Bessel functions,
and the approximation holds when $\mu>>1$. We use SI units in this Appendix. The effect of finite
length is small, provided that length $\ell \geq 2b$, which is the
case for the experiment. In the case $a=b$ for a loop directly on a
cylinder, $\delta L\approx 6\mu_0 a\approx 0.1\ \mu$H, for $a=1.2$
cm, a 20\% effect.  When $a=\sqrt{2} b$, $\delta L=0.02\ \mu$H, a
much smaller effect.  The loss in sensitivity is around 10\%,
leaving the principal problem due to this effect as the detuning of
the gradiometer condition, which is temperature dependent, and of
order 10\%.  Usually, gradiometers are fabricated to provide 1\%
common mode rejection.  Thus, we see a substantial limitation to the
degree of common mode rejection possible when there is a permeable
sample in the central loop.  In principle, the loop areas could be
adjusted to cancel gradients at a particular temperature, but
introduces difficulties if temperature variation is employed to test
for systematic effects.

\bibliography{References}

\end{document}